\documentclass[a4paper, oneside, twocolumn, notitlepage, 10pt]{extarticle_ecoc}
\usepackage{ecoc}
\usepackage{tikz}
\usepackage{pgfplots}
\usepackage{TUMcolors}
\usepackage[nolist]{acronym}

\begin{document}
\selectlanguage{english}

\begin{acronym}
    \acro{FEC}{forward error correction}
    \acro{oFEC}{open FEC}
    \acro{SO-SCL}{soft-output successive cancellation list}
    \acro{SOGRAND}{soft-output guessing random additive noise decoding}
    \acro{SOCS}{soft-output from covered space}
    \acro{eBCH}{extended Bose-Chaudhuri-Hocquenghem}
    \acro{BPSK}{binary phase shift keying}
    \acro{AWGN}{additive white Gaussian noise}
    \acro{LLR}{log-likelihood ratio}
    \acro{APP}{a posteriori probability}
    \acro{BER}{bit error rate}
    \acro{SNR}{signal-to-noise ratio}
    \acro{ML}{maximum likelihood}
\end{acronym}


\title{Decoding Product Codes and Staircase Codes with Iteration-Independent Weighting Coefficients}%


\author{
    Andreas Straßhofer\textsuperscript{(1)}
}

\maketitle                  


\begin{strip}
    \begin{author_descr}

        \textsuperscript{(1)} School of Computation, Information and Technology, Technical University of Munich, Germany, 
        \textcolor{blue}{\uline{andreas.strasshofer@tum.de}}

    \end{author_descr}
\end{strip}

\renewcommand\footnotemark{}
\renewcommand\footnoterule{}


\begin{strip}
    \begin{ecoc_abstract}
        This paper presents an improved FEC decoder design outperforming Chase-Pyndiah decoding of product codes by $0.23$ dB. To achieve this, the decoder does not require iteration-dependent coefficients, making it implementation-friendly for sliding-window decoding of staircase codes. ©2026 The Author
    \end{ecoc_abstract}
\end{strip}


\section{Introduction}

Optical communication systems require energy-efficient decoders for \ac{FEC} codes to support data rates of up to $400$ Gbps. At the same time, small additional coding gains can significantly expand the reach of a link, motivating the search for near-optimal decoders.

Product codes have become a testbed for decoders, often benchmarked against the Chase-Pyndiah decoder \cite{Pyndiah1998}. This decoder is list-based; however, it struggles to produce meaningful soft output when the list size is small. Recent results from \ac{SO-SCL} decoding \cite{Peihong2025} and \ac{SOGRAND} \cite{Peihong2025_2} employ novel soft-output computation. The underlying principle was later applied to product codes in \cite{Janz2025}. The herein-proposed \ac{SOCS} decoder achieves a gain of $0.25$ dB over Chase-Pyndiah decoding; however, it still uses iteration-dependent weighting coefficients and requires computations in the probability domain. In this paper, we build upon these works and propose a novel decoder without iteration-dependent weighting coefficients supporting log-domain implementation. Despite this reduction in complexity, the results show no loss in performance for product codes. Another significant advantage of the proposed decoder is its ease of use and good performance for sliding-window decoding of spatially coupled product-like codes, such as staircase codes.

The paper is structured as follows. We start by introducing preliminaries, followed by an outline of the Chase-Pyndiah decoder. We then show the proposed decoder and outline its advantages over Chase-Pyndiah decoding. In the final sections, we present numerical results for product and staircase codes, followed by a conclusion.

\section{Preliminaries}
In this work, we consider product codes of which each row and each column is protected by an \ac{eBCH} constituent code $\mathcal{C}$ of length and dimension $(n, k)$. The code rate is $k^2/n^2$.

Staircase codes are chains of $n/2 \times n/2$ blocks $\mathbf{B}_i$ for $i \geq 0$ for which $\mathbf{B}_0$ is a matrix of known bits. Herein the constituent code protects each row of $[\mathbf{B}_{i-1}^\text{T},\mathbf{B}_i]$. The code rate of unterminated staircase codes is $2 \frac{k}{n} - 1$.

We use \ac{BPSK} and map bits $c_i$ to channel inputs $x_i$ as $0 \rightarrow +1$ and $1 \rightarrow -1$. The channel is memoryless and subject to \ac{AWGN}. We therefore have $y_i = x_i + n_i$ where $n_i \sim \mathcal{N}(0,\sigma^2)$. The decoder operates on channel \acp{LLR}
\begin{align}
    l^\text{ch}_i = \ln \left( \frac{p_{\mathsf{Y} \mid \mathsf{X}}(y_i \mid +1)}{p_{\mathsf{Y} \mid \mathsf{X}}(y_i \mid -1)} \right)
\end{align}
which evaluate to $l^\text{ch}_i = \frac{2}{\sigma^2} \cdot y_i$ for the \ac{AWGN} channel with inputs $\left\{+1,-1\right\}$.

During iterative decoding, each channel \ac{LLR} is combined with a priori information from previous iterations, giving the vector $\boldsymbol{l}$ of inputs to the constituent decoder. The optimal soft-output is the logarithmic \ac{APP} ratio, which for bit $i$ is
\begin{align} \label{eq:optimal}
    l_i^{\mathrm{app}} = \ln \left( \frac{\displaystyle\sum_{\boldsymbol{c} \in \mathcal{C}
    :c_i = 0} P_{\mathbf{C} \mid \mathbf{L}} (\boldsymbol{c} \mid \boldsymbol{l})}%
    {\displaystyle\sum_{\boldsymbol{c} \in \mathcal{C}:c_i = 1} P_{\mathbf{C} \mid \mathbf{L}} (\boldsymbol{c} \mid
    \boldsymbol{l})} \right)\,.
\end{align}
The exact extrinsic information is $\boldsymbol{l}^{\mathrm{e}} = \boldsymbol{l}^{\mathrm{app}} - \boldsymbol{l}$. Using Bayes' law and the memoryless channel each summand can be written as
\begin{align} \label{eq:Bayes}
	P_{\mathbf{C} \mid \mathbf{L}} (\boldsymbol{c} \mid \boldsymbol{l}) = \frac{P_\mathbf{C} (\boldsymbol{c})}{p_\mathbf{L} (\boldsymbol{l})} \cdot \prod_{i=1}^{n} \frac{p_\mathsf{L} (l_i)}{P_\mathsf{C}(c_i)} P_{\mathsf{C} \mid \mathsf{L}} (c_i|l_i)\,.
\end{align}

Computing \eqref{eq:optimal} is prohibitively complex for practically relevant constituent codes, such as two-error-correcting \ac{eBCH} codes.

\section{Chase-Pyndiah Decoding of Product Codes}

\begin{figure*}[t]
    \centering
    \includegraphics[scale=0.7]{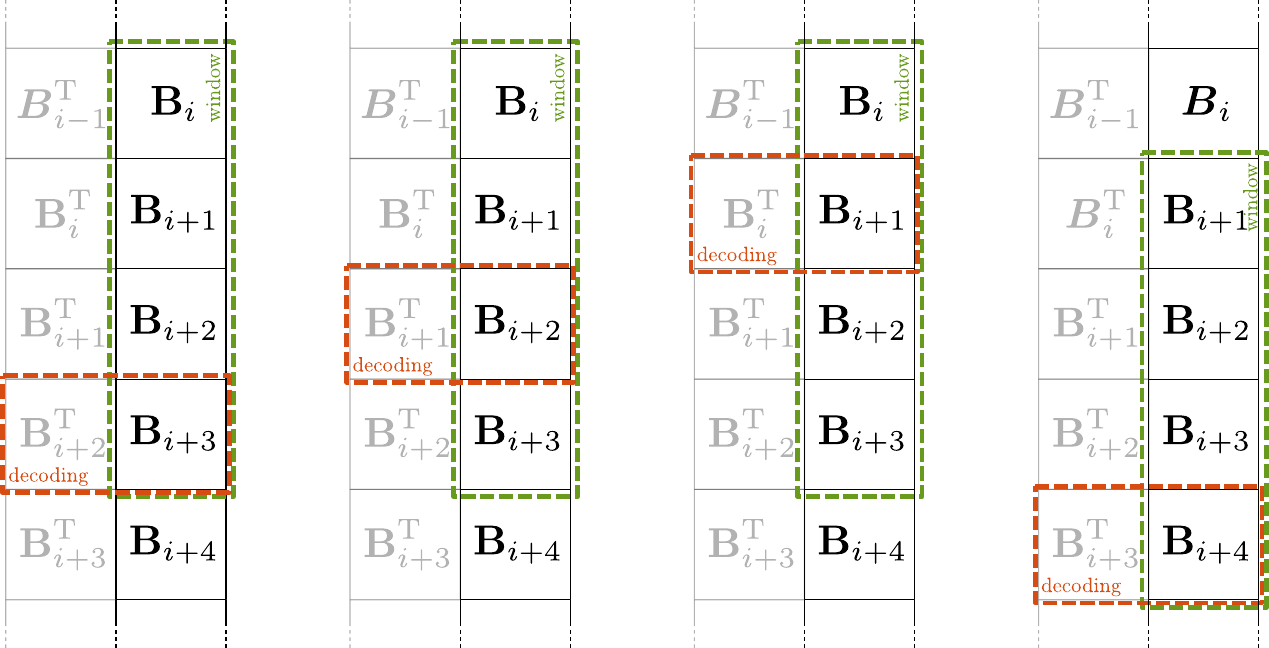}
    \vspace{4mm}
    \caption{Sliding window decoding of a staircase code with window size $w=4$. Already decided blocks are italic.}
    \label{fig:sliding_window_decoding}
\end{figure*}

The Chase-Pyndiah decoder computes a soft output based on the following two approximations.
\vspace{0mm}
\begin{enumerate}
    \item Instead of the whole codebook $\mathcal{C}$ we use a list $\mathcal{L}$ found by Chase-II decoding of the constituent code.
    \item We approximate each sum in \eqref{eq:optimal} by its maximum summand, known as the max-approximation.
\end{enumerate}
For uniformly distributed and independent information bits, and a linear code the approximate logarithmic \ac{APP} ratio becomes
\begin{align} \label{eq:Chase_Pyndiah_start}
    \widetilde{l_i^\mathrm{app}} = \frac{1}{2} \left( \max_{\boldsymbol{c} \in \mathcal{L}:c_i = 0}
    \langle \boldsymbol{x}, \boldsymbol{l} \rangle -
    \max_{\boldsymbol{c} \in \mathcal{L}:c_i = 1}
    \langle \boldsymbol{x}, \boldsymbol{l} \rangle \right)
\end{align}
where $\langle \cdot , \cdot \rangle$ denotes the inner product. If the sets $\left\{ \boldsymbol{c} \in \mathcal{L}:c_i = 0 \right\}$ and $\left\{ \boldsymbol{c} \in \mathcal{L}:c_i = 1 \right\}$ are not empty we compute the approximate extrinsic information as
\begin{align} \label{eq:normal}
    \widetilde{l_i^\mathrm{e}} = \alpha^{(\ell)} \cdot \left( \frac{1}{|\mathcal{J}|}\sum_{j \in \mathcal{J}} \mid  \widetilde{l_j^{\mathrm{app}}} - l_j \mid  \right)^{-1} \left( \widetilde{l_i^{\mathrm{app}}} - l_i \right)
\end{align}
where $\alpha^{(\ell)}$ is the weighting coefficient in half-iteration $\ell$. The set $\mathcal{J}$ contains all indices of the whole product codeword for which the aforementioned sets are not empty. Otherwise, the extrinsic information is
\begin{align} \label{eq:alternative}
    \widetilde{l_i^\mathrm{e}} = \alpha^{(\ell)} \cdot \left( \frac{1}{|\mathcal{J}|}\sum_{j \in \mathcal{J}} \mid  \widetilde{l_j^{\mathrm{app}}} - l_j \mid  \right)^{-1} \beta^{(\ell)} \cdot x_i
\end{align}
where $\beta^{(\ell)}$ is another iteration-dependent weighting coefficient and $x_i$ is the modulated codebit on which all codewords in $\mathcal{L}$ agree upon. The inputs for the following half-iteration are then found as 
\begin{align} \label{eq:Chase_Pyndiah_end}
    l_i = \widetilde{l_i^\mathrm{e}} +  \left( \frac{1}{n^2}\sum_j \mid  l_j^{\mathrm{ch}} \mid \right)^{-1} l_i^\mathrm{ch} \,.
\end{align}
In the last half-iteration of Chase-Pyndiah decoding, we choose, for each row (or column), the \ac{ML} codeword from the respective list as the final decision.

\section{Proposed Decoder}

We propose to approximate the logarithmic \ac{APP} ratio as
\begin{align} \label{eq:proposed}
    \widetilde{l_i^{\mathrm{app}}} = \ln \left( \frac{\displaystyle\sum_{\boldsymbol{c} \in \mathcal{L}
    :c_i = 0} P_{\mathbf{C} \mid \mathbf{L}} (\boldsymbol{c} \mid \boldsymbol{l}) + \gamma \cdot P_{\mathsf{C} \mid \mathsf{L}} (0 \mid l_i)}
    {\displaystyle\sum_{\boldsymbol{c} \in \mathcal{L}:c_i = 1} P_{\mathbf{C} \mid \mathbf{L}} (\boldsymbol{c} \mid
    \boldsymbol{l})  + \gamma \cdot P_{\mathsf{C} \mid \mathsf{L}} (1 \mid l_i) } \right)
\end{align}
where $\gamma$ is an iteration-independent weighting coefficient. We then compute the extrinsic information as $\widetilde{\boldsymbol{l}^{\mathrm{e}}} = \widetilde{\boldsymbol{l}^{\mathrm{app}}} - \boldsymbol{l}$. Our proposed decoder has the following key advantages over Chase-Pyndiah decoding.
\vspace{4mm}
\begin{itemize}
    \item Reduction of complexity by the use of a single iteration-independent weighting coefficient.
    \item No matter the list $\mathcal{L}$, the numerator and denominator are non-zero, eliminating the need for \eqref{eq:alternative}.
    \item Incorporation of all codewords in the list as the max-approximation is reversed, bringing the form of the proposed decoder closer to the optimal soft-output in \eqref{eq:optimal}.
\end{itemize}

It remains to efficiently compute $P_{\mathbf{C} \mid \mathbf{L}} (\boldsymbol{c} \mid \boldsymbol{l})$. We therefore approximate \eqref{eq:Bayes} as
\begin{align} \label{eq:random_coding}
    P_{\mathbf{C} \mid \mathbf{L}} (\boldsymbol{c} \mid \boldsymbol{l}) &\approx \prod_{i=1}^{n} P_{\mathsf{C} \mid \mathsf{L}} (c_i|l_i) \nonumber \\ 
    &= \exp \left( - \sum_{i=1}^{n} \ln \left( 1 + \exp ( - x_i l_i ) \right) \right)
\end{align}

where the sum is known as the path metric, a quantity readily available if $\mathcal{L}$ was obtained by successive cancellation list decoding \cite{Tal2015}. We remark that \eqref{eq:random_coding} holds with equality under a random-coding argument \cite{Peihong2025_2} and under iterative decoding over a cycle-less graph; neither is applicable to product codes or staircase codes.

\section{Application to Staircase Codes}

We decode staircase codes using a sliding window spanning $w$ blocks. For each window position, we start by decoding the first $n/2$ constituent codes. These protect the block that entered the window most recently. The extrinsic information is used as a~priori to decode the next $n/2$ constituent codes. This procedure continues $w-1$ times.

The final decisions for the last block in the window are obtained from decoding the $n/2$ second-to-last constituent codes. As in Chase-Pyndiah decoding, we take the bits of the \ac{ML} codeword over the list. We then shift the decoding window by one block and restart the procedure while conserving any a priori information. Fig.~\ref{fig:sliding_window_decoding} visualizes the procedure for $w=4$ using the zipper code framework from \cite{Sukmadji2022}.

We remark that the same weighting coefficient $\gamma$ is used throughout the decoding procedure, i.e., for decoding every constituent code within the window.

\section{Simulation Results}

\begin{figure}[t]
    \centering
    \begin{tikzpicture}
        \begin{semilogyaxis}[
            width=\columnwidth,
            height=7cm,
            xmin=3,
            xmax=4.1,
            xtick distance=0.1,
            xlabel={$E_\text{b}/N_0$ [dB]},
            ymin=1e-7,
            ymax=1e-1,
            ylabel={Bit Error Rate},
            legend pos=south west,
            legend cell align=left,
            legend style={font=\footnotesize},
            ticklabel style={font=\footnotesize},
            label style={font=\footnotesize},
            grid=both,
            grid style={opacity=0.5},
            ]
            
            \addplot[TUMpantone300, mark=*, mark options={fill=white}, line width=1.5] table [x=Eb_N0, y=BER] {CP.txt};
            \addlegendentry{Chase-Pyndiah \cite{Pyndiah1998}};
            
            \addplot[TUMgold, mark=*, mark options={fill=white}, line width=1.5] table [x=Eb_N0, y=BER] {SOCS_B_L.txt};
            \addlegendentry{SOCS($\mathcal{B}_2(\mathcal{L})$) \cite{Janz2025}};
            
            \addplot[TUMlightgreen, mark=*, mark options={fill=white}, line width=1.5] table [x=Eb_N0, y=BER] {proposed.txt};
            \addlegendentry{Proposed $\gamma = 2^{-17}$};
            
        \end{semilogyaxis}
    \end{tikzpicture}
    \caption{BER performance for a rate $0.872$ product code with $(256,239)$ eBCH constituent codes.}
    \label{plot:product_codes}
\end{figure}
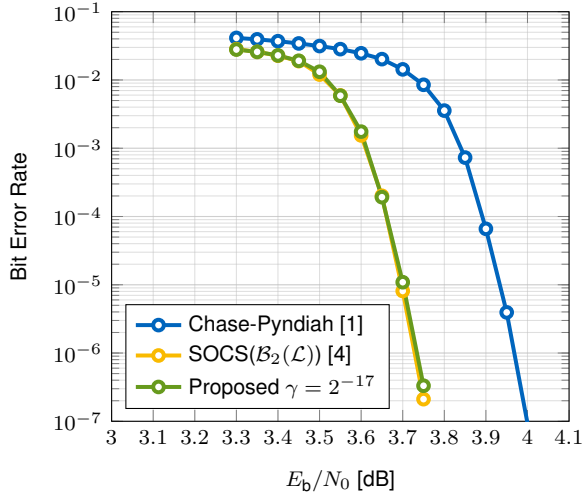

We first evaluate the \ac{BER} performance of the proposed decoder on product codes. We compare to Chase-Pyndiah decoding and the recently proposed \ac{SOCS} decoder. The constituent code is a $(256,239)$ \ac{eBCH} code. Each decoder obtains a list from Chase-II decoding by evaluating $2^5 = 32$ error patterns. We run iterative decoding for $4$ iterations ($8$ half-iterations) and the channel \acp{LLR} are from an \ac{AWGN} channel with \ac{SNR} $E_\text{b}/N_0 = \left( 2R\sigma^2 \right)^{-1}$ where $R$ is the code rate. The results are depicted in Fig.~\ref{plot:product_codes}.

The proposed decoder outperforms Chase-Pyndiah decoding by around $0.23$ dB at a \ac{BER} of $10^{-6}$. Despite its reduced complexity, achieved by using a single weighting coefficient $\gamma = 2^{-17}$ and no probability-domain computations, we obtain results similar to those of the best \ac{SOCS} decoder from \cite{Janz2025}.

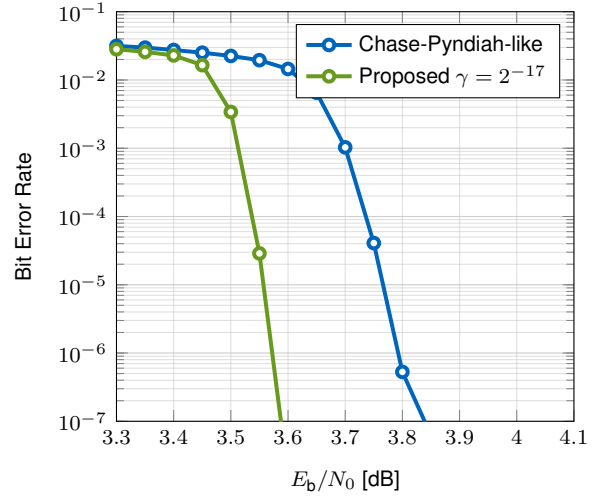
\begin{figure}[t]
	\centering
	\begin{tikzpicture}
		\begin{semilogyaxis}[
            width=\columnwidth,
            height=7cm,
            xmin=3.3,
            xmax=4.1,
            xtick distance=0.1,
            xlabel={$E_\text{b}/N_0$ [dB]},
            ymin=1e-7,
            ymax=1e-1,
            ylabel={Bit Error Rate},
            legend pos=north east,
            legend cell align=left,
            legend style={font=\footnotesize},
            ticklabel style={font=\footnotesize},
            label style={font=\footnotesize},
            grid=both,
            grid style={opacity=0.5},
            ]
            
            \addplot[TUMpantone300, mark=*, mark options={fill=white}, line width=1.5] table [x=Eb_N0, y=BER] {SC_Chase_Pyndiah_like.txt};
			\addlegendentry{Chase-Pyndiah-like};

            \addplot[TUMlightgreen, mark=*, mark options={fill=white}, line width=1.5] table [x=Eb_N0, y=BER] {SC_proposed.txt};
			\addlegendentry{Proposed $\gamma = 2^{-17}$};
			
		\end{semilogyaxis}
	\end{tikzpicture}
    \caption{BER performance for a rate $0.867$ staircase code with $(256,239)$ eBCH constituent codes.}
    \label{plot:staircase_codes}
\end{figure}

Next, we evaluate the proposed decoder on staircase codes using sliding window decoding with a window size of $w = 8$. The counting of bit errors starts after the first $20$ blocks, and at least $10^5$ bit errors are collected. We compare to a Chase-Pyndiah-like decoder, which uses equations (\ref{eq:Chase_Pyndiah_start}-\ref{eq:Chase_Pyndiah_end}) with the following modifications. We reverse the max-approximation in \eqref{eq:Chase_Pyndiah_start} and omit the normalization terms in (\ref{eq:normal}-\ref{eq:Chase_Pyndiah_end}). We set $\alpha = 0.4$ and $\beta = 3.6$, which are found via a grid search with a step size of $0.1$. The results are depicted in Fig.~\ref{plot:staircase_codes}.

The proposed decoder gains around $0.15$ dB compared to product decoding and outperforms the Chase-Pyndiah-like decoder by $0.22$ dB, both at a \ac{BER} of $10^{-6}$. To make a fair comparison with product codes, we may apply an \ac{SNR} penalty of about $0.05$ dB to account for the slightly lower rate of the staircase code.

\section{Conclusions}

We proposed the approximate \ac{APP} decoding rule \eqref{eq:proposed} which avoids the iteration-dependent weighting coefficients present in Chase-Pyndiah decoding and the \ac{SOCS} decoder. For product codes, we outperform Chase-Pyndiah decoding by $0.23$ dB and match \ac{SOCS} decoding with the advantage that the proposed decoder can be implemented entirely in the log domain. Results for an example staircase code show gains of $0.22$ dB over a Chase-Pyndiah-like approach. 

\section{Acknowledgement}

The author thanks Tim Janz and Simon \mbox{Obermüller} for insightful discussions on SOCS decoding.

\printbibliography

\vspace{-4mm}

\end{document}